\documentclass[twocolumn,showpacs,aps,floatfix]{revtex4} 
\usepackage{graphicx}
\newcommand{\ignore}[1]{}
\def\vareps{\varepsilon}

\begin{document}
%
\title{
Electron transmission and phase time in semiconductor superlattices} 
\author{D. W. L. Sprung$^{a}$, W. van Dijk$^{a,b}$ 
and C. N. Veenstra$^{a,b,c}$}
\affiliation{$^a$Department of Physics and Astronomy, McMaster University, 
Hamilton ON, L8S 4M1}
\affiliation{$^b$Redeemer University College, Ancaster ON L9K 1J4, Canada} 
\address{$^c$Department of Physics \& Astronomy, University of 
Britush Columbia, Vancouver BC, V6T 1Z1} 
\author{J. Martorell }
\affiliation{Departament d'Estructura i Constituents de la Materia, Facultat
F\'{\i}sica, University of Barcelona\\  Barcelona 08028, Spain}
\date{\today}
%
\begin{abstract}
We discuss the time spent by an electron propagating through a finite 
periodic system such as a semiconductor superlattice. The relation 
between dwell-time and phase-time is outlined. The envelopes of 
phase-time at maximum and minimum transmission are derived, and it is 
shown that the peaks and valleys of phase-time can be well described 
by parameters fitted at the extrema. For a many-period system this 
covers most of the allowed band. Comparison is 
made to direct numerical solutions of the time-dependent 
Schr\"odinger equation by Veenstra et al. \{cond-mat/0411118\}  who 
compared systems with and without addition of an anti-reflection 
coating (ARC). With an ARC, the time delay is consistent with 
propagation at the Bloch velocity of the periodic system, which 
significantly reduces the time delay, in addition to increasing the 
transmissivity. 
 \end{abstract}
\maketitle
\section{Introduction} 
The question of  ``how long does it take for a non-relativistic 
particle to cross a barrier", or in general any potential, is a 
contentious one in quantum physics \cite{Hau89,Leav96,Raz03}. 
There is for example the Hartmann effect \cite{Hart62}, according to which a 
transmitted particle might be found before it has reached the 
potential. Since 1990 the subject has taken on renewed interest in 
the context of electrons in semiconductor superlattices (SL)
\cite{Nuss00,CN02,IP94,I95,DSDA96,GR97,CS98,P00,SP03,PBG05,ML00,muga07}. 
It is commonly assumed that 
such propagation is ballistic, with a mean free path larger than the 
device dimensions. But because electrons can scatter from lattice 
vibrations, it is of interest to know how long the electron is exposed to 
such interactions. Many theories have been put forward as to how the 
time of passage should be defined and measured; among them dwell-time and 
phase-time are the best recognized.  Veenstra et al. \cite{veenstra} 
studied time dependence of propagation by direct numerical solution 
of the time-dependent Schr\"odinger equation (TDSE), using gaussian 
incident wave packets. Their results agreed well with phase-time. In 
this contribution, based on a talk given at Theory-Canada 3 in June 2007, 
we explain why phase-time should be applicable in 
the context of semiconductor superlattices.

\section{Transfer Matrix method}

Assuming ballistic transport, a conduction band electron in a 
potential cell of arbitrary shape on $a < x < b$ satisfies 
the Schr\"odinger equation with a material-dependent effective mass:
 \begin{eqnarray}
&& m^*(x) \frac{d}{dx} \left[\frac{\,\,\,d \psi(x)}{m^*(x)d x}\right] 
+ \frac{2m^*(x)}{\hbar^2} [E - V(x)] \psi(x) 
 = 0 \nonumber \\
&& \quad \,\,\,\, k^2(x) = \frac{2m^*(x)}{\hbar^2} [E - V(x)] \nonumber \\
&& \quad \Psi(x,t) = e^{-iEt/\hbar} \,\,\psi(x)~.
\label{eq:tc01} 
\end{eqnarray}
The approximations leading to eq.\ref{eq:tc01} are fully explained 
in Bastard's monograph \cite{GB88}. 
At a layer boundary, $\psi(x)$ and $\psi^\prime \equiv (\hbar/m^*) 
d\psi/dx$ are continuous, to conserve flux. It is convenient to match 
$\psi(x)$ to plane waves, normalized to unit flux, at the boundaries 
of the unit cell $a<x<b$ \cite{FPP} 
 \begin{eqnarray}
\psi_L(x) &=& \frac{c_L}{\sqrt{v_L}} e^{ik_L(x-a)} + 
\frac{d_L}{\sqrt{v_L}}  e^{-ik_L(x-a)}  \quad , \, x < a ,
\nonumber \\ 
\psi_R(x) &=&  \frac{c_R}{\sqrt{v_R}} e^{ik_R(x-b)} + 
\frac{d_R}{\sqrt{v_R}}  e^{-ik_R(x-b)}   \quad , \, x > b~. 
\label{eq:tc02} 
\end{eqnarray}
At either end, ($B = L,\, R$) $k_B$ is the wave number outside the 
periodic system, and $v_B = \hbar k_B/m_B^*$ is the velocity. We will 
consider systems without bias, so $k_L = k_R = k$, etc., but it will 
be convenient to retain the indices in order to know where various 
terms come from. The transfer matrix relates the coefficients on 
either side: 
 \begin{eqnarray}
\pmatrix{c_L \cr d_L} &=& \pmatrix{M_{11} & M_{12} \cr M_{21} & M_{22} }
                      \pmatrix{c_R \cr d_R }
\quad {\rm where} \nonumber \\ 
M &=& \pmatrix{1/t & r^*/t^* \cr r/t & 1/t^*} 
\label{eq:tc03}
\end{eqnarray} 
in terms of the reflection and transmission amplitudes $r(k)$, $t(k)$. 
It contains all the information about scattering from that cell. 
One can show \cite{SMM04} that 
 \begin{eqnarray}
M \sigma_z M^\dagger &=& \sigma_z \quad  
\Rightarrow   |c|^2 - |d|^2 = {\rm constant}~. 
\label{eq:tc04} 
\end{eqnarray}
which says that the probability flux is preserved by the action of 
$M$, and which implies $ \det M = 1$. 

\begin{figure}[htb]                    
\begin{center}
\begin{tabular}{cc} 
a) & b) \\
\includegraphics[width=4cm]{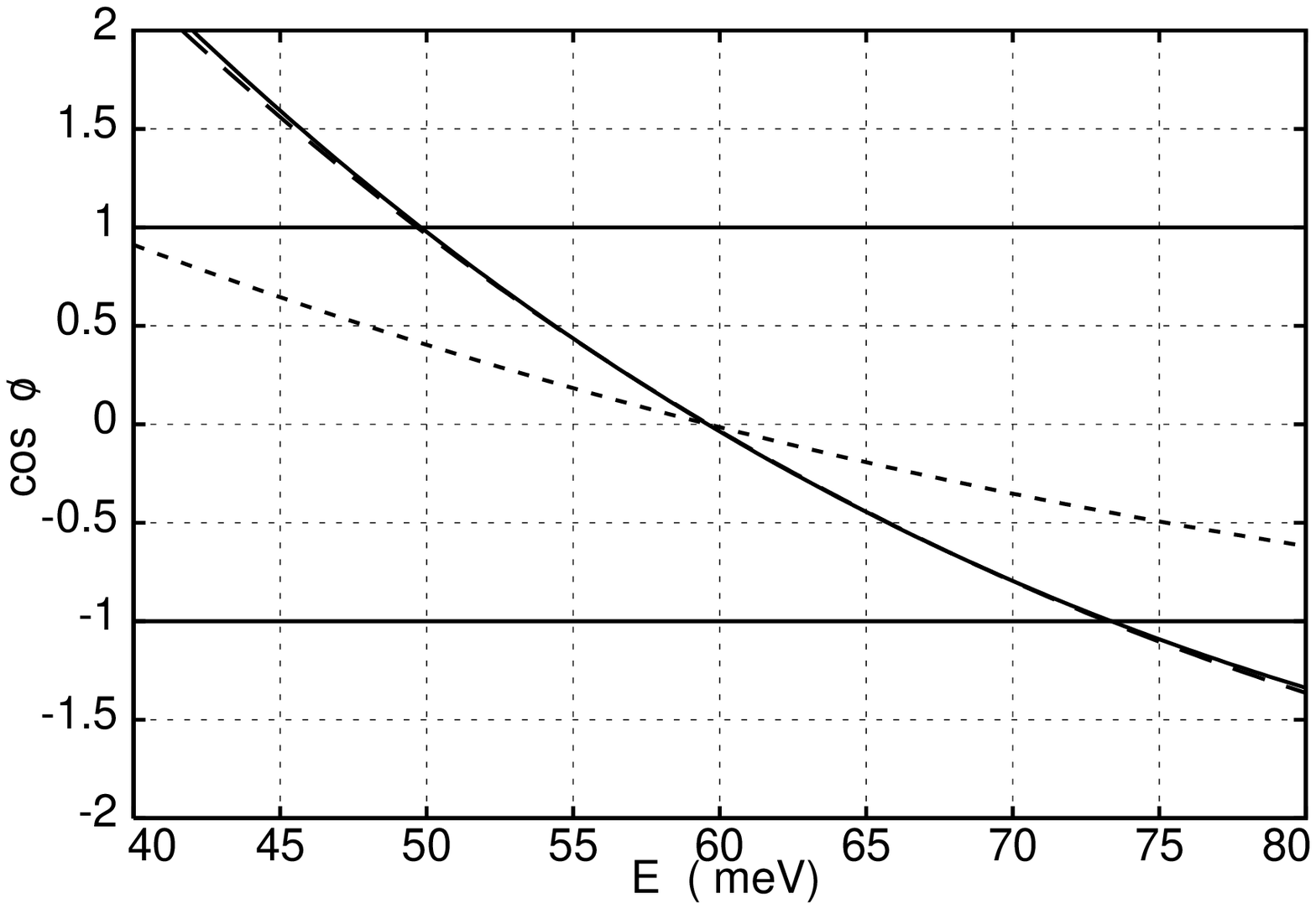} & 
\includegraphics[width=4cm]{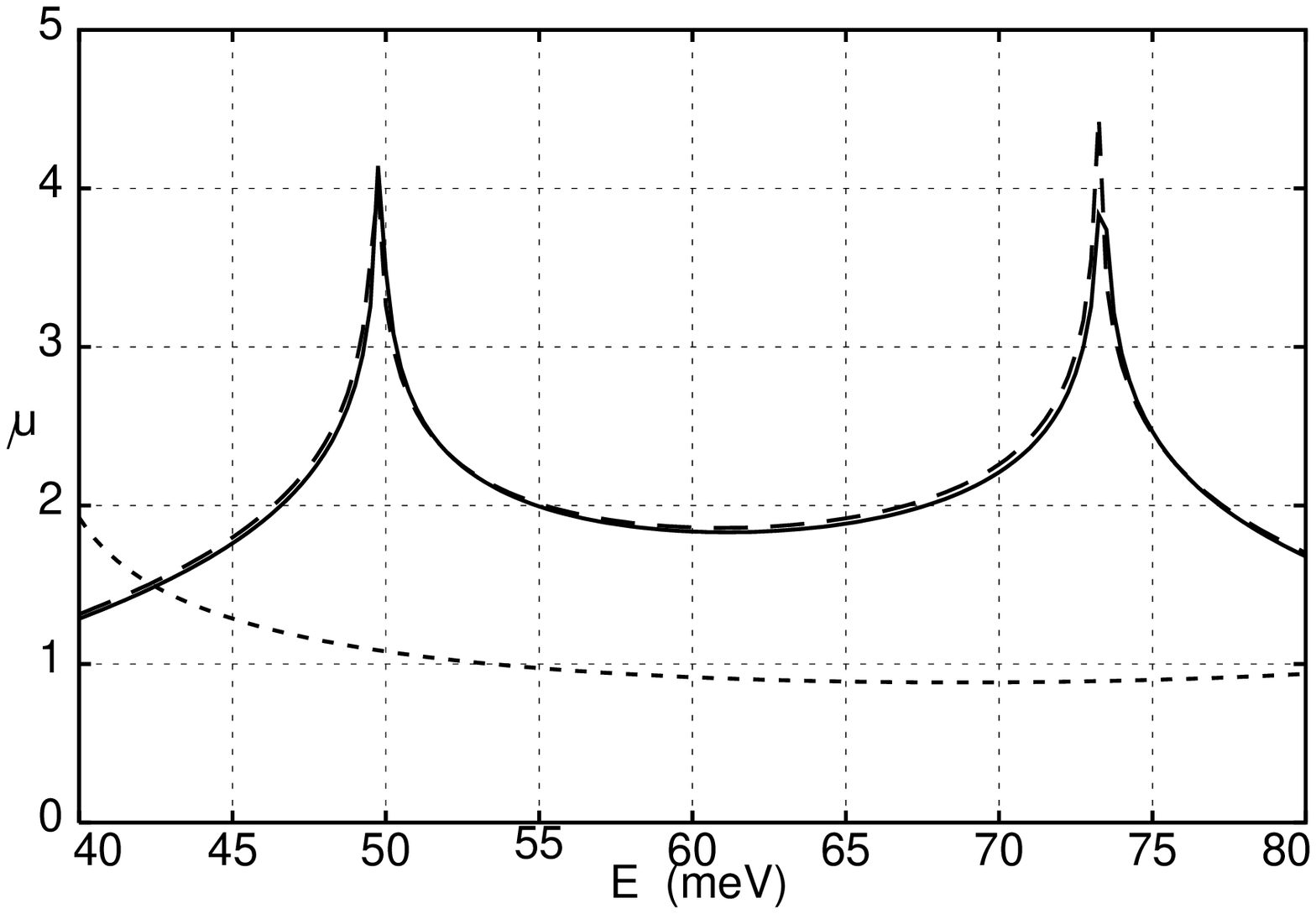} \\ 
\end{tabular}
\end{center}
\caption{Comparison of (a) Bloch phases of a square barrier cell 
(dash line) and a gaussian cell (solid line); also shown is 
$\cos \phi_A$ of a single-cell ARC (dotted line). 
(b) same for impedance parameters $\mu, \, \mu_A$.} 
\label{fig0}
\end{figure}

It is convenient to use Kard's parameterization \cite{Kard57} of $M$. In an 
allowed band we write 
 \begin{eqnarray}    
M_{11}  &=& \cos \phi -i \sin \phi \,\cosh \mu = 1/t = M_{22}^* 
\nonumber \\
M_{21}  &=& \phantom{\cos \phi} -i \,e^{i\chi}\, \sin \phi\, \sinh \mu 
=  r/t = M_{12}^* \, .
\label{eq:tc05} 
\end{eqnarray}
At a given energy, the Bloch phase is determined by 
Tr$M = 2\cos \phi$; the impedance parameter from the 
ratio of $|M_{21}|/$Im$M_{11} = \tanh \mu$, and the asymmetry 
parameter $\chi$ from the  phase of $M_{21}/M_{12}= \exp{(2i\chi)}$. 
($\chi = 0 $ for a reflection-symmetric potential cells. For the most 
part this paper is restricted to the symmetric case.) 
For a simple square barrier-cell, $e^\mu$ is the ratio of average velocity 
outside to inside. Across an allowed band, $\phi$ increases by $\pi$, 
while $\mu$ is quite constant across most of the band, diverging at the 
band edges. 
An example of this behaviour is shown in Fig. \ref{fig0}. taken from 
\cite{veenstra}. Panel (a) 
show $\cos \phi$ for a square barrier cell, and for a phase-shift 
equivalent gaussian barrier cell. Panel (b) shows the corresponding 
values for $\mu$. The dotted lines in each panel are the parameters 
of the single-layer ARC cell. The slope of $\cos \phi_A$ is about 
half that of $\cos \phi$, while their Bragg points coincide. The rule 
of thumb for ARC's in optics would make $\mu_A = \mu/2$.  

In the Kard representation, transfer matrix factorizes: 
 \begin{eqnarray}
&& M = 
\cos \phi  
 - i\sin \phi \bigl[ \cosh \mu + \sinh \mu 
\bigl(\cos \chi \sigma_x + \sin \chi \sigma_y \bigr) \Bigr] \sigma_z 
\nonumber \\ 
&&  = e^{-i(\chi/2) \sigma_z} \,\, e^{(\mu/2) \sigma_x} \,\, e^{-
i\phi \sigma_z} \,\, e^{-(\mu/2) \sigma_x} \,\, e^{i(\chi/2) \sigma_z} 
\label{eq:tc06}
\end{eqnarray}
The matrix of eigenstates is easily seen to be 
$$U = e^{-i(\chi/2) \sigma_z} e^{(\mu/2) \sigma_x}$$ 
Substitution into eq. \ref{eq:tc06} gives
 \begin{eqnarray}
M U  &=& U \,\, e^{-i \phi \sigma_z} \, ,  \nonumber \\ 
U &=& \pmatrix{ e^{-i \chi/2} \cosh \mu/2 &  e^{-i \chi/2} \sinh \mu/2 \cr 
\,\, e^{i \chi/2} \sinh \mu/2 &\,\,  e^{i \chi/2} \cosh \mu/2 }  
\label{eq:tc07} 
\end{eqnarray}
The eigenvalues are $e^{- i \phi}$ and $e^{+ i \phi}$. 
Physically, the eigenstates are the Bloch waves of an infinite 
periodic array composed of cells, each described by $M$. 
Mathematically, $M$ is a hyperbolic rotation of a two-dimensional 
Dirac spinor, about an invariant axis whose polar angles are $(\mu,\, 
\chi)$ \cite{SMM04}. 

The usefulness of the transfer matrix for a periodic system 
arises from the property 
 \begin{eqnarray}
M^{(N)} &=& M^N(\phi, \mu, \chi) = M(N\phi, \mu, \chi) \quad \Rightarrow 
\nonumber \\ 
\frac{1}{t_N} &=& \cos N\phi - i \sin N\phi \cosh \mu~. 
\label{eq:tc08}
\end{eqnarray}
Hence the transmission probability for $N$-cells 
 \begin{eqnarray}
|t_N|^2 &=& [1 + \sin^2 N\phi \, \sinh^2 \mu]^{-1} \ge 1/\cosh^2 \mu 
\label{eq:tc09}
\end{eqnarray}
shows narrow peaks determined by $N\phi(E) = m \pi$, $m = 1, 2, 
\cdots N-1$. Between these peaks the $|t_N|^2$ touches the envelope 
of minima, $1/\cosh^2 \mu$. 
At the center of an allowed band, the envelope of minima crosses the curve 
$|t_1|^2$.                   

In forbidden bands, the Bloch phase $\phi\to p\pi + i\theta$ acquires an 
imaginary part; as a result the transmission goes rapidly to zero.  
Pacher et al. \cite{Pach01} used this property to design an electron 
band-pass filter. Further, by adding a quarter-wave cell 
(anti-reflection coating, or ARC) at each end of the periodic array 
\cite{PG03,PG04}, they were able to increase the average transmission 
within the band from about $25\%$ to about $75\%$.  
In a series of papers \cite{MSM02,MSM02b,SMM03,SMM04}, some of us 
showed how to design an ARC which gives optimal transmission within a 
given miniband, by adding suitably configured potential cells on each 
end of a periodic array. 
Without an ARC, electrons which are transmitted do so via narrow 
resonances as in eq. \ref{eq:tc09}, so one expects a significant time 
delay. With an ARC, the incident plane wave is transformed into a 
Bloch wave of the periodic potential, and travels at the Bloch 
velocity. This should reduce the transit time.

\section{Time Delay} 
\subsection{Dwell time and phase time}
As already mentioned, the subject of time-delay in scattering is 
controversial. Razavy's book \cite{Raz03} is a useful introduction. 
Nussenzweig \cite{Nuss00,CN02} has argued that dwell-time is the best 
founded, for description of wave packet scattering. 
First we explain two of these theories. 

Classically, 
 \begin{equation}
dt = {{dx}\over {v(x)}} = {{dx} \over {\hbar k(x)/m}} \ . 
\label{eq:tc10}
\end{equation}
For a plane wave normalized to unit flux, as in eq. \ref{eq:tc02}, 
 \begin{equation}
|\psi_{0}(x,E)|^2 = {m^* \over{\hbar k}} = \frac{1}{v_{cl}}~. 
 \label{eq:tc11}
\end{equation}
Dwell-time delay is defined to be the difference 
 \begin{equation}
\tau_D = \int_{x_L}^{x_R} \left[ |\psi(x,E)|^2 - 
|\psi_0(x,E)|^2\right] \ dx ~.
 \label{eq:tc12}
\end{equation}
Asymptotically, the scattering wave function, for waves incident from 
the left, is usually written 
 \begin{eqnarray}
\psi(x,E) &=& \frac{1}{\sqrt{v_L}} 
\left[ e^{i k_L x} + r\, e^{-ik_L x}\right] \quad , \quad x <x_L \nonumber \\ 
 &=& \frac{1}{\sqrt{v_R}} t\, e^{i k_R x} \qquad , \qquad x > x_R~, 
\nonumber \\ 
{\rm where} && t = |t| e^{i\eta} \quad , \quad r = |r| e^{i\delta}
\label{eq:tc13}
\end{eqnarray}
define the scattering phase shifts. 

Following Smith \cite{S60}, the Schr\"odinger equation gives the 
identity 
 \begin{eqnarray}
&& \,\, \psi = (H-E) \frac{\partial \psi}{\partial E} \qquad \Rightarrow 
\nonumber \\ 
&&\,\, \psi^* \psi = - \frac{\hbar^2}{2m} {\rm Re} \left\{ 
\frac{\partial}{\partial x} \left[ \psi^* \frac{\partial}{\partial x} 
\left(\frac{\partial \psi}{\partial E}\right) -  
\frac{\partial \psi}{\partial E} \frac{\partial \psi^*}{\partial x}  
\right]\right\}~, \nonumber \\ 
&&\int_{x_L}^{x_R} \, \psi^* \psi dx =
- \frac{\hbar^2}{2m} {\rm Re} \left[ \left.\psi^* \frac{\partial}{\partial x} 
\left(\frac{\partial \psi}{\partial E}\right) - 
\frac{\partial \psi}{\partial E} \frac{\partial \psi^*}{\partial x}  
\right]\,\, \right|_{x_L}^{x_R}\nonumber \\
\label{eq:tc14}
\end{eqnarray} 
For a potential on $a < x <b$, $x_L < a$ and $x_R > b$ are positions 
at which a position measurement could be carried out. 
Taking only the scattering wave function in eq. \ref{eq:tc12}, 
 \begin{eqnarray}
-{\hbar^2\over{2m}} {\rm Re} \left[\psi^* {{\partial}\over{\partial 
x}}\left( {{\partial \psi}\over{\partial E}} \right) -{{\partial 
\psi}\over{\partial E}} {{\partial \psi^*}\over{\partial x}} 
\right]_{x_R} &=&  \left({x_R\over{v_R}} + \hbar {{\partial 
\eta}\over{\partial E}}\right) |t|^2 \nonumber \\ 
 -{\hbar^2\over{2m}} {\rm Re} \left[\psi^* {{\partial}\over{\partial 
x}}\left( {{\partial \psi}\over{\partial E}} \right) -{{\partial 
\psi}\over{\partial E}} {{\partial \psi^*}\over{\partial x}} 
\right]_{x_L} &=&   \nonumber \\ 
{x_L \over {v_L}} ( 1 + |r|^2 )  
- \hbar {{\partial \delta}\over {\partial E} } |r|^2 
+ {m |r| \over{\hbar k_L^2}} && \sin(2 k_L x_L -\delta)~. \nonumber \\ 
\label{eq:tc15}
\end{eqnarray} 
Using eqs. \ref{eq:tc15} and \ref{eq:tc14} in eq. \ref{eq:tc12} gives 
the dwell-time as 
 \begin{eqnarray} 
\tau_{D} &=& 
\hbar\left({{\partial \eta}\over{\partial E}} |t|^2 
+ {{\partial \delta}\over{\partial E}} |r|^2 \right) -{{\hbar |r|}\over 
{2 E}} \sin(2k_L x_L -\delta) \nonumber \\ 
&& + 
\left({x_R\over v_R}-{x_L\over v_L}\right) |t|^2 
-  \left({2x_L\over v_L}\right) |r|^2 ~.
 \label{eq:tc16}
\end{eqnarray} 
Since the potential lies between the limits, we can assume that $x_L$ 
is a negative distance, while $x_R$ is positive. The terms on the 
second line have an obvious interpretation as the ``free passage 
time", so the top line is the ``dwell-time delay". But be careful: 
Nussenzweig \cite{Nuss00} has a lucid discussion of the case of an 
incident wave-packet, and finds an additional effect which arises 
from the uncertainty principle: you cannot localize a quantum 
particle in less than a de Broglie wave length. This effect cancels 
out between the free and interacting situations, so it affects 
dwell-time delay, but not dwell-time. We  skip over this complication 
and rely on eq. \ref{eq:tc16}. 

Phase time was introduced by Wigner and Eisenbud \cite{W54}, and 
similarly relates time delay to the rate of change of the scattering 
phase. Scattering in 1D is a two-channel problem (incident waves from 
left, or right). The $S$-matrix is symmetric and unitary, and may be 
written 
 \begin{equation}
S = \pmatrix{ r & t \cr t & \bar{r}} = 
\pmatrix{ |r|e^{i\delta} & |t| e^{i\eta} \cr |t| e^{i\eta} & 
-|r|e^{i(2\eta -\delta)} } 
 \label{eq:tc17}
\end{equation}
where $\bar{r} = -r^* t/t^*$ is the reflection amplitude for 
waves incident from the right. $|r|^2 + |t|^2 = 1$. 

Smith \cite{S60} defined the time-delay matrix for a many-channel 
system as 
 \begin{eqnarray}
&& \quad \tau \equiv -i\hbar S^\dagger \frac{dS}{dE} = 
-i\hbar S^\dagger \,\, \times \nonumber \\ 
&&\pmatrix{(|r|^\prime + i |r| \delta^\prime ) e^{i\delta} & 
(|t|^\prime + i |t| \eta^\prime) e^{i\eta} \cr 
(|t|^\prime + i |t| \eta^\prime) e^{i\eta} & 
- (|r|^\prime + i |r| (2\eta^\prime -  \delta^\prime)) e^{i(2\eta-\delta)} } 
\nonumber \\ 
\label{eq:tc18}
\end{eqnarray}
where primes mean derivative with respect to energy. 
After some work we have 
 \begin{eqnarray}
\tau_{11} &=& 
+ \hbar\,\, \left( |t|^2 \frac{d\eta}{dE} + |r|^2 \frac{d\delta}{dE} \right) 
\nonumber \\ 
\tau_{12} &=& -i \hbar e^{i(\eta - \delta)}  \frac{\partial}{\partial E} 
\left( \tan^{-1} (\frac{|r|}{|t|}) \right) +  \hbar r^* t 
\left( \frac{d\eta}{dE} - \frac{d\delta}{dE} \right)
\nonumber \\ 
\tau_{22} &=& 
+ \hbar\,\, \left( \frac{d\eta}{dE} + |r|^2 
\left( \frac{d\eta}{dE} - \frac{d\delta}{dE} \right)
\right)~.
 \label{eq:tc19}
\end{eqnarray}

In the case of a reflection symmetric potential, $\bar{r} = r$, which 
requires $\eta = \delta + \pi/2$, so their derivatives are equal. As 
a result, all elements $\tau_{ij}$ are real, and the diagonal 
elements are equal. 
For a reflection-symmetric potential we have the compact result 
 \begin{eqnarray}
\tau_{11} &=&  \hbar\, \frac{d\eta}{dE} = \tau_{22} 
\nonumber \\ 
\tau_{12} &=&  \hbar \frac{\partial}{\partial E} 
\left( \sin^{-1} {|r|} \right) = \hbar ({|r|^\prime}/{|t|})~.
 \label{eq:tc20}
\end{eqnarray}
The diagonal elements are called the ``phase-time delay". They agree with 
the first term in the top line of eq. \ref{eq:tc16}. But due to 
coupling, the time associated with a process depends on just which 
mixture of the two channels is involved. The difference is that eq. 
\ref{eq:tc16} was derived assuming that one has specified the channel 
with incident waves from the left, ignoring the other channel. 
The oscillatory term of eq. \ref{eq:tc16} has been subject of much 
debate also. Obviously it arises from interference between the 
incident and reflected waves. Winful \cite{Win03} intepreted this 
term as the time taken to cross a distance of the order of the 
scattering length.  On a more practical note, if $x_L \to -\infty$, 
and one averages $k_L$ over a wave packet narrow in energy, this 
oscillatory term is exponentially small. Therefore in a practical 
sense, either dwell-time or phase-time are equivalent for SL 
scattering, at least for symmetric potentials.

Eq. \ref{eq:tc20} shows that there are two situations where the 
channel coupling vanishes: at a maximum or minimum of transmission. 
In a SL, the number of maxima is $N-1$, so as $N$ increases, these 
points of uncoupling become closer together.

\subsection{A tale of two phases}

In the transfer matrix method, the wave functions are defined with a 
different phase than in usual scattering theory:  we reset the phase 
to zero on each side ($x=a,\, b$) of the potential array, rather than 
at an arbitrary origin. That makes $M$ translation invariant, so $M^N$ 
describes a periodic system without having to include a shift in the 
origin. 

Adopting Nussenzweig's notation $[\psi_L\,; \, \psi_R]$ for the 
asymptotic wave function to left and right \cite{Nuss00},  we compare 
our $\psi(x)$ with the usual convention $\tilde{\psi}(x)$ as follows: 
 \begin{eqnarray}
\psi(x) &\sim& [ e^{ik(x-a)} + r e^{-ik(x-a)} \, ; \, t e^{ik(x-b)} ] 
\nonumber \\  
\tilde{\psi}(x) &\sim& [ e^{ikx} + \tilde{r} e^{-ikx} \, ; \qquad 
\quad \tilde{t} e^{ikx} \quad ] 
\nonumber \\  
e^{ika}\, \psi(x) &\sim& [ e^{ikx} + r e^{-ik(x-2a)} \, ; \,\,\,\,\, 
\,\, t e^{ik(x-w)} 
]~, 
\label{eq:tc21}
\end{eqnarray} 
where $w = b-a$ is the total width of the potential. It follows that 
the phase $\eta$ of {\it our} transmission amplitude is related to the 
usual phase by $\tilde{\eta} = \eta - kw$. Then the standard phase-time 
delay is  
 \begin{eqnarray}
\tau_{ph} = \hbar \frac{d \tilde{\eta}}{d E} 
 &=& \hbar \Bigl[ \frac{d \eta}{d E} - w \frac{d k}{d E} \Bigr] 
= \hbar \Bigl[ \frac{d \eta}{d E} -  \frac{w}{v_{cl}} \Bigr]~.  
\label{eq:tc22}
\end{eqnarray} 
We conclude that the phase $\eta$ of our transfer matrix amplitudes 
gives phase time, not time-delay, because the free-passage time has 
to be subtracted from it.

\section{Phase time for superlattice transmission}
\subsection{Relation of scattering phases to Kard parameters} 

The $S$-matrix and the transfer matrix contain the scattering 
information in different forms. To see how phase-time applies to a 
SL, we examine the relation between the two descriptions. Equating 
 \begin{equation}
{1\over t_N} =  \cos N\phi - i \sin N \phi \, \cosh \mu
\equiv {1\over |t_N|}\, e^{-i\eta_N}
 \label{eq:tc23}
\end{equation}
we find the following relations between the two sets of parameters: 
 \begin{eqnarray}
\cos \eta_N  = |t_N| \, \cos N \phi  \, &;& \quad 
\sin \eta_N  = |t_N| \, \sin N \phi \cosh \mu  \nonumber \\ 
\tan \eta_N = \tan N\phi \cosh \mu \, &;& \quad 
\frac{\tan \eta_N}{\tan N\phi} = \cosh \mu = \frac{\tan \eta}{\tan 
\phi}~, \nonumber \\ 
 \label{eq:tc24}
\end{eqnarray} 
where $\eta = \eta_1$ is the phase shift for a single cell. 
In the first allowed band, $0 < \phi < \pi$; $\tan \eta$ varies 
smoothly; $\cosh \mu$ diverges at the band edge, like $1/\sin \phi$. 
Zeroes and poles of $\tan \eta_N$ coincide with those of $\tan 
N\phi$. These points are the scattering resonances $N\phi_m = m\pi$ 
and the minima of transmission $N\phi_p = (p + 1/2)\pi$. 
Near the poles, $\eta_N$ lags behind $N\phi$, then it has to catch up 
at the zeroes $N \phi_m$. This is nicely illustrated in Fig. \ref{fig4}. 
The steeper slope of $/\eta_N$ near $\phi_m$ 
makes for a longer time delay at those energies.  

\subsection{\bf Phase time near the maxima of transmission} 
In an infinite periodic array the electron would move at the Bloch 
velocity. Let the time to cross a cell of width $d$ be $\tau_{Bl}$. 
Write $\phi/d = \kappa$, the pseudo-momentum.  
 \begin{eqnarray} 
\frac{\partial E}{\partial \phi} =  \frac{1}{d} 
\frac{\partial E}{\partial  \kappa} &=& \frac{\hbar^2 \kappa}{d\, m^* } 
= \frac{\hbar}{d} v_{Bl} = \frac{\hbar}{\tau_{Bl}} \nonumber \\ 
\hbar \frac{\partial \phi}{\partial E} &=& \tau_{Bl}~. 
\label{eq:tc25} 
\end{eqnarray}  

In the following, the prime will mean $\partial/\partial E$. 
We can show using eqs. \ref{eq:tc23} and \ref{eq:tc24} that 
 \begin{eqnarray}
\frac{\partial \eta_N}{\partial E} &=&  
{\left[ N\phi^\prime \,\cosh \mu + \sin 2N\phi \,\, \sinh \mu\,\, 
\mu^\prime/2 \right] 
\over  \left[1 + \sinh^2 \mu \,\,\sin^2 N\phi\right] } \nonumber \\ 
\tau_{ph} &=& N \tau_{Bl} \cosh \mu \,\, {
\left[ 1 + \sin 2N\phi \,\, \tanh \mu\,\, (\mu^\prime/(2 N 
\phi^\prime) \right] 
\over  \left[1 + \sinh^2 \mu \,\,\sin^2 N\phi\right] } \nonumber \\  
\label{eq:tc26}
\end{eqnarray}

Away from resonance, the denominator is a 
factor of $|t_N|^2$ which cuts off the phase time 
very sharply, causing it to mimic the shape of the transmission 
curve. Near a resonance, $\phi \sim m\pi/N + \vareps$, so that $$(-
)^m \sin N\phi \sim \sin N \vareps \sim N (E - E_m) \phi_m^\prime.$$ 
This allows us to approximate $|t_N|^2$ as a Breit-Wigner resonance, with 
half-width $\Gamma_m = 2/(N \sinh \mu_m \,\, \phi_m^\prime)$: 
 \begin{eqnarray}  
|t_N|_m^{2} = \left[ 1 + (\frac{E - E_m}{\Gamma_m/2 })^2 \right]^{-1}~. 
\label{eq:tc27}
\end{eqnarray} 
Similarly, in the same vicinity, 
 \begin{eqnarray}
&&\tau_{ph, m} = \left. \hbar \frac{\partial \eta_N}{\partial E} \right|_{m} 
\sim  N \tau_{Bl,m} \cosh \mu_m \times \nonumber \\ 
&&\,\,\qquad \times {\left[ 1 + 2 b_m 
(\frac{E - E_m}{\Gamma_m/2 })  \right] {\Big/}  
\left[1 + (\frac{E - E_m}{\Gamma_m/2 })^2 \right]}  
\nonumber \\ 
&&{\rm with} \,\,\, 2\, b_m =   \frac{1}{N \phi_m^\prime \, \cosh \mu_m} 
\left[ 2\mu_m^\prime + \coth \mu_m \frac{\phi_m^{\prime\prime}}
{\phi_m^{\prime}} \right]\nonumber \\ 
\label{eq:tc28}
\end{eqnarray}
This is called a Fano resonance shape \cite{Rau04}. From eq. 
\ref{eq:tc26}, we see that the locus of phase time at transmission 
maxima is 
 \begin{eqnarray}
\tau_{ph,{\rm max}} = N \tau_{Bl} \, \cosh \mu 
\label{eq:tc29}
\end{eqnarray}
Similarly, at transmission minima, $\sin^2 2N\phi =  1$; the 
denominator of $\tau_{ph}$ (eq. \ref{eq:tc26}) becomes $\cosh^2 \mu$, 
giving a downside locus of 
 \begin{eqnarray}
\tau_{ph,{\rm min}} = N \tau_{Bl}/\cosh \mu 
\label{eq:tc30}
\end{eqnarray}
for phase time at transmission minima. 
The Bloch time for $N$ cells is the geometric mean of the two loci.

\subsection{Phase time near the minima}

Transmission minima occur at 
$\phi_p = (p+1/2) \pi/N$, for $p = 1,\, 2,\, \cdots n-2.$ 
Close by, the denominator may be  written 
 \begin{eqnarray}
&& t_N^{-2} = \cosh^2 \mu ( 1 - \tanh^2 \mu\,\, \cos^2 N\phi )
\nonumber \\ 
&& \sim \cosh^2 \mu_p \left[ 1 + \mu_p^\prime \tanh \mu_p (\delta E_p) 
\right]^2 \,  \left[ 1 - ( N \phi_p^\prime\, \tanh \mu_p \, \delta E_p)^2 
\right] \nonumber \\ 
&& \,\, = \cosh^2 \mu_p \left[ 1 + \frac{\mu_p^\prime}{N \phi_p^\prime} 
\left( \frac{\delta E_p}{\Gamma_p/2}\right) \right]^2 \, 
\, \left[ 1 - \left( \frac{\delta E_p}{\Gamma_p/2}\right)^2 \right]
\label{eq:tc31}
\end{eqnarray}
where $\delta E_p = E - E_p$. The width 
 \begin{eqnarray}
\Gamma_p = 2 / [N \phi_p^\prime \tanh \mu_p]
\label{eq:tc32}
\end{eqnarray}
is large compared to the widths $\Gamma_m$ of the resonances. 

 \begin{eqnarray}
&&\tau_{ph, p} = \hbar \frac{\partial \eta_N}{\partial E}\Bigr|_{p} 
\nonumber \\
&& \sim   \frac{N \tau_{Bl,p} }{\cosh \mu_p} \frac{ \left[ 1 + C_p 
\left( \frac{E - E_p}{\Gamma_p/2} \right)  \right]} 
{\left[ 1 + 2 D_p \left(\frac{E - E_p}{\Gamma_p/2}  \right)  
+ (D_p^2 -1)\, \left(\frac{E - E_p}{\Gamma_p/2 }\right)^2 \right]}   
\nonumber \\
\label{eq:tc33}
\end{eqnarray}
 \begin{eqnarray}
&& {\rm where} \,
C_p  =   \frac{\phi_p^{\prime\prime} }{\phi_p^{\prime}} \, 
\frac{\Gamma_p}{2} \qquad {\rm and} \qquad 
D_p = \frac{\mu_p^{\prime} }{N \phi_p^{\prime}}~. 
\label{eq:tc34}
\end{eqnarray}
Both $C_p$ and $D_p$ are of order $1/N$. 
The prefactor is the locus of time-delay at minima: 
 \begin{eqnarray}
\frac{N \tau_{Bl} }{\cosh \mu} = \frac{N \hbar }{\cosh \mu} 
\frac{d \phi}{d E} 
= - N \hbar \frac{d \cos \phi}{d E}\, /\, {\rm Im} {M_{11}} 
\label{eq:tc35}
\end{eqnarray}
which involves only well-behaved single-cell quantities.

Veenstra et al. \cite {veenstra} compared their computed time 
delay to phase-time delay, and found good agreement. In particular 
they found that the locus of maxima and eqs. \ref{eq:tc27}, 
\ref{eq:tc28} accounted for the overall picture. In an 
interesting paper, Pacher, Boxleitner and Gornik \cite{PBG05} pointed 
out that all theories of time-delay agree at the transmission maxima, 
so they concentrated their attention on those points and found the 
locus of maxima. The main difference between their work and the 
present one is that they spoke of the mean velocity for an 
electron traversing the SL, rather than the time. Further they 
expressed their results in terms of the real and imaginary parts of the 
transfer matrix elements $M_{ij}$, e.g. eq. \ref{eq:tc35}, 
rather than the Kard parameters. 
It is our opinion that the Kard parameters, given their simple 
behaviour,  make the expressions more easily understandable.

\subsection{\bf Play Model}
To illustrate the above results, we take a simple model in which 
we specify $\cos \phi$ to be linear in energy in a band between $E= 
50$ and $75$ meV. In addition, $|t|^{2}$ is specified. All other 
parameters follow from these two. With such a model it is easy to see 
how changing some parameter will affect the results. 
 \begin{eqnarray}
\cos \phi &=& 0.08(62.5 - E)\, ,  \quad 50 \le E \le 75\,\, ({\rm meV}) 
, \nonumber \\ 
|t|^{-2} &=&  1 + 160/E ~. 
\label{eq:tc36}
\end{eqnarray}
The phase $\eta$ is determined by $\cos \eta = |t| \cos \phi$, and 
the impedance parameter follows from 
 \begin{eqnarray}
\cosh \mu &=& \sin \eta / (|t| \sin \phi)~. 
\label{eq:tc37}
\end{eqnarray}
It diverges at each band edge, since $|t|\,\sin \eta $ is a smooth function of 
energy, while $\phi$ runs from $p \pi$ to $(p+1) \pi$.

\begin{figure}                 
\includegraphics[width=8cm]{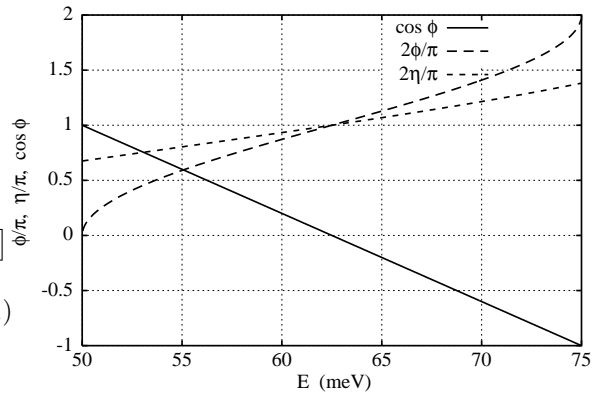}
\caption{ Play model: $\cos \phi$, and angles $\phi$, and $\eta$, in 
units of $\pi/2$. Remarkably, $\phi$ is reasonably linear over much 
of the band.}  
 \label{fig1} 
\end{figure}

In Fig. \ref{fig1} 
we show the play model $\cos \phi = \lambda(E_B - E)$, and the 
corresponding phases $\phi$ and $\eta$. $\eta$ is quite linear, 
crossing $\phi$ at the Bragg point. In this model, $\phi^\prime = 
\lambda/\sin \phi$, with $\lambda = 0.08$ meV$^{-1}$. 

 \begin{figure}                 
\includegraphics[width=8cm]{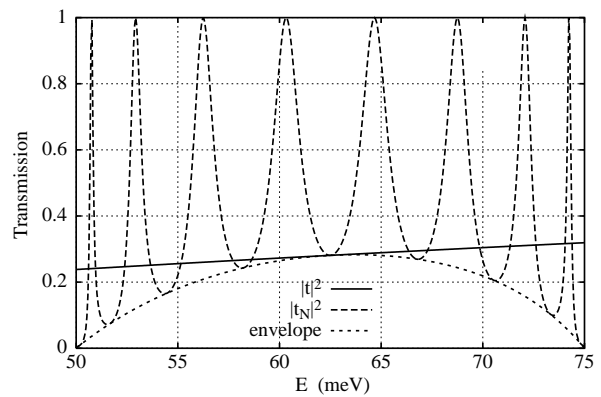}
\caption{ Play model: Transmission probability for one cell, nine cells, 
and envelope of transmission minima.} 
 \label{fig2} 
\end{figure}

In Fig. \ref{fig2} 
we show the play model transmission for 9 cells. 
The envelope of transmission minima is simply $1/ \cosh^2 \mu$, which 
gives direct physical significance to the impedance parameter.

 \begin{figure}                 
\includegraphics[width=8cm]{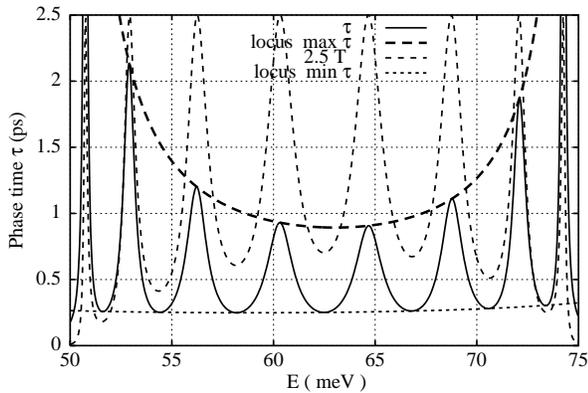}
\caption{ Play model: phase time, and loci of phase time at 
maxima and minima of transmission. Transmission probability is shown 
for orientation. } 
 \label{fig3} 
\end{figure} 

 \begin{figure}[htb]                 
\begin{center}
\includegraphics[width=8cm]{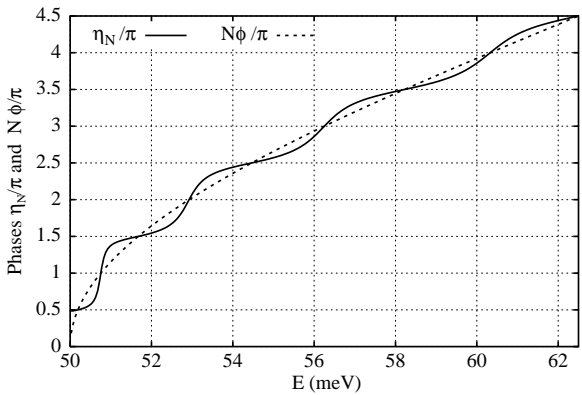}
\end{center}
\caption{ Play model: $\eta_N$ and $N \phi$, in units of $\pi$. 
Note that slope of $\eta_N$ is largest at integer multiples of $\pi$.
For clarity, only half the allowed band is shown.} 
 \label{fig4} 
\end{figure} 

 \begin{figure}                 
\includegraphics[width=8cm]{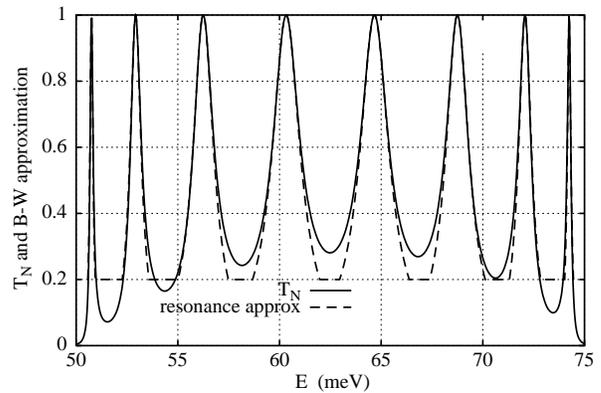}
\caption{ Transmission probability compared to resonance 
approximation (eq. \protect{\ref{eq:tc27}}). }
 \label{fig5} 
\end{figure} 

In Fig. \ref{fig3} 
we show the play model phase-time, along with the envelopes of maxima 
and minima. In the background for orientation are the transmission 
peaks, which line up well with the maxima of phase-time. 

In Fig. \ref{fig4} 
we show the dance of the Bloch phase for $N$-cells, and the 
corresponding transmission phase $\eta_N$, in the lower half of the 
allowed band. The curve in the upper half of the band is a double 
reflection of this, ending at $9\pi$. 
Except at the band edges, the lines cross at every 
half-integer multiple of $\pi$. Since $\eta_N$ is catching up at 
integer multiples of $\pi$, the steeper slope leads to a longer 
phase-time at the transmission resonances. 

In Fig. \ref{fig5} 
we show (solid line) the transmission and (dashed line) the 
Breit-Wigner resonance fitted at the peaks. This is truncated at two 
standard deviations; the horizontal lines simply connect the B-W curves 
between successive peaks. The agreement is excellent. 

\begin{figure}                 
\includegraphics[width=8cm]{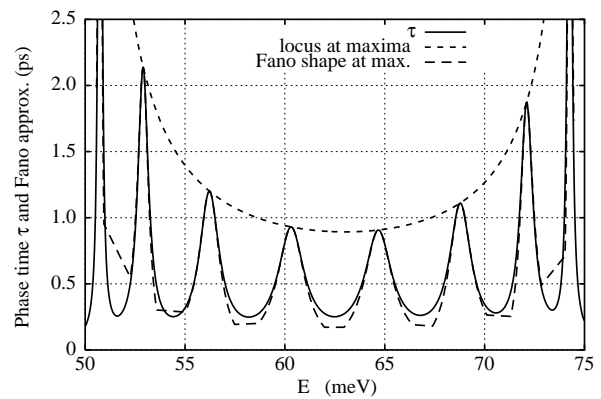}
\caption{ Play model: phase time, and its approximation by resonance 
formula eq. \protect{\ref{eq:tc28}}. Locus of phase time at maximum 
transmission is shown to demonstrate that phase-time maxima are very little 
shifted from the locus.} 
 \label{fig6} 
\end{figure}

Fig. \ref{fig6} 
is similar to Fig. \ref{fig5}, but for the phase-time. The 
Fano-shape formula fitted to the maxima also does an excellent job of 
reproducing the exact calculation. 

\subsection{Realistic potential model}
\begin{figure}                 
\includegraphics[width=8cm]{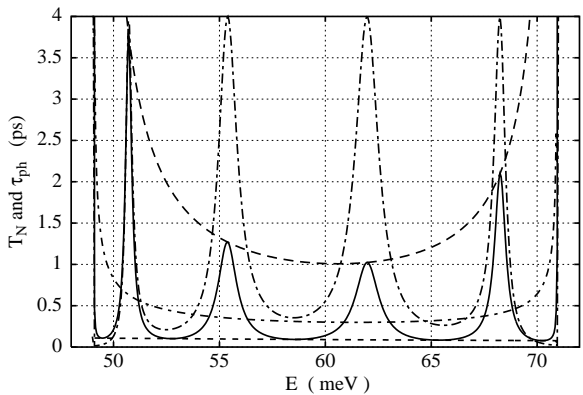} 
\caption{ Pacher-Gornik 5-cell array: phase time (solid line),  along 
with the loci at maxima and minima (long dashes), and Bloch time 
(dash - dot), which is their geometric mean. The chain-line is four 
times the transmission probability.} 
\label{fig7} 
\end{figure}

\begin{figure}                 
\includegraphics[width=8cm]{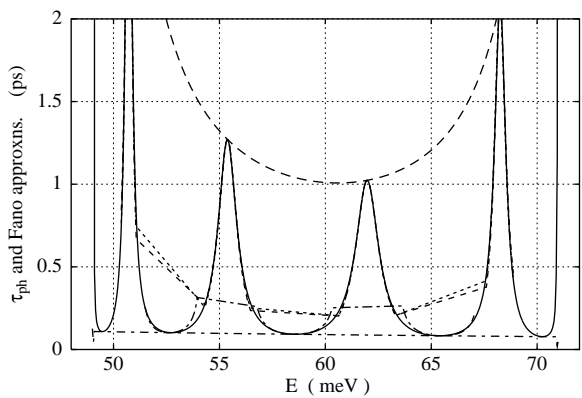}
\caption{ Pacher-Gornik 5-cell array: along with the 
Fano-type approximation at each resonant 
peak and valley. The fit at the peaks is excellent over two 
half-widths, but not so good at the phase-time minima.
The long dash and dash-dot lines are the loci of phase-time at 
maximum and minimum transmission.} 
 \label{fig8} 
\end{figure}

Some results using a realistic semiconductor potential of Pacher and 
Gornik \cite{PG03,PG04} are shown in Figs. \ref{fig7}, \ref{fig8}. 
Here we took five potential cells, giving four resonances in the 
band, a number chosen for comparison with the calculations of 
Veenstra et al. \cite{veenstra}.
Fig. \ref{fig7} corresponds to Fig. \ref{fig3} for the phase-time. 
For this potential the envelope of minima is lowermeaning a larger 
$\mu$. Also shown is the 
Bloch time, the geometric mean of the two envelopes. 

Fig. \ref{fig8} can be compared with Fig. \ref{fig6}. The Fano formula fitted 
at the maxima of transmission agrees with the exact result over two 
standard deviations. In Fig. \ref{fig8} we have included eq. 
\ref{eq:tc33}, fitted at the minima of transmission. The fit to the 
minima is not so good, largely because the half-width at minima is so 
much larger than at maxima. Still, we can say that  between them the 
approximations reproduce the phase time over about 90\% of the band. 
If there were more layers, the peaks and valleys would be narrower 
and the agreement would improve, as in Fig. \ref{fig6}. For large 
$N$ the phase time can be calculated reliably using only properties 
at the extrema, where the various theories of time-delay agree with 
each other.

Finally, in Fig. \ref{fig9} 
we show the results calculated by Veenstra et al. for the Pacher five 
cell array, plus a two layer ARC. The dotted line shows the 
transmission is close to 100\% over most of the band width. There are two 
lines shown for the Bloch time. One takes into account only the 
periodic 5-cell system, and the other includes a correction for the 
ARC layers on each end. The two estimates are close together. The time delay 
extracted from the TDSE calculations is indeed very close to the 
Bloch time, which is what we expect from the argument that the ARC 
converts the incident plane wave into a Bloch state.

\section{Conclusion} 
We have outlined the transfer matrix method for transmission of 
electrons in a one-dimensional AlGaAs/GaAs superlattice, using the 
Kard parameterization. We then compared the two most commonly used 
theories for the average time spent in crossing a potential region, 
namely phase time and dwell time. These were applied to SL 
transmission, first for a play model and then for the potential that 
corresponds to experiments of Pacher and Gornik. We noted that 
the phase-time is well defined both at maxima and minima of 
transmission. In the neighbourhood of these points, the transmission 
can be described by a Breit-Wigner resonance formula, with 
the parameters extracted at the extrema. The same holds for the 
phase-time, except it is a Fano shape resonance. 
The fits at maxima are good over two half-widths, and at 
the minima over one. Taken together, this covers almost all the band 
width, for a system with more than a very few periods. 

As the number of periods $N$ of the SL is increased, the widths of 
the peaks and valleys decrease, and the approximate forms become more 
and more accurate. This explains the success of 
phase-time for describing the time spent in traversing a SL.

 \begin{figure}                 
\begin{center}
\includegraphics[width=7cm]{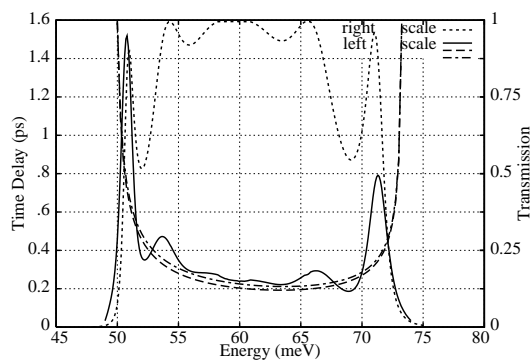}
\end{center}
\caption{Time-dependent numerical solution of wave equation for a 
Pacher-Gornik 5-cell array plus ARC:  phase-time compared to Bloch 
time. Transmission shown in background.} 
 \label{fig9} 
\end{figure} 

The locus of phase-time at maxima was derived in Veenstra et al. 
\cite{veenstra}, and by Pacher et al. \cite{PBG05}. The locus of 
phase-time at mimina is new. Their geometric average is the Bloch time 
for traversing a SL. An ARC works by converting the incident plane 
wave into a Bloch wave of the periodic system. Veenstra et al. 
\cite{veenstra} studied the time delay for the SL plus ARC by direct 
numerical solution of the TDSE using gaussian wave packets. In Fig. 
\ref{fig9}, taken from that reference, it can be seen that indeed 
the time-delay with ARC agrees quite well with the Bloch time. 
The ARC not only increases the average transmission, but it smooths 
out the dwell time, removing the peaks and valleys associated with 
the exponential decay of the resonances.

\acknowledgements 

We are grateful to NSERC-Canada for Discovery Grants SAPIN-8672 
(WvD), RGPIN-3198 (DWLS) and a Summer Research Award through Redeemer 
University College (CNV); and to DGES-Spain for continued support 
through grant FIS2006-10268-C03-01 (JM). We also 
thank Gigi Wong for assistance in redrawing Fig. \ref{fig9}.



\end{document}